# Plated-Through-Hole Via Design Specifications for 112G Serial Links

Michael J. Degerstrom, Mayo Clinic
degerstrom.michael@mayo.edu

Chad M. Smutzer, Mayo Clinic
smutzer.chad@mayo.edu

Richard B. Ericson, Mayo Clinic
ericson.richard@mayo.edu

Clifton R Haider, Mayo Clinic
haider.clifton@mayo.edu

Barry K. Gilbert, Mayo Clinic
gilbert.barry@mayo.edu, 507-284-4056

## Abstract

An earlier study of a high layer-count test board using plated-through-hole (PTH) vias and a limited quantity of laser vias was shown to be capable of supporting 112 Gb/s PAM-4 links (or equivalent signaling having 28 GHz (Nyquist) bandwidth). This original board design was then rebuilt using a different fabricator, and the test results revealed a significant decrease in the bandwidth of the vias. These results led to the development of a set of design specifications that PCB vendors can easily validate, which will ensure that the use of high layer-count boards with PTH technology are viable for emerging 112 Gb/s PAM-4 links.

## Biographies

**Michael J. Degerstrom** received a BSEE from the University of Minnesota. Mike is currently a Senior Engineer at the Mayo Clinic Special Purpose Processor Development Group. His primary area of research and design has been in the specialty of signal and power integrity.

**Chad M. Smutzer** received a BSEE from the University of Iowa in Iowa City. He is currently a Senior Engineer at the Mayo Clinic Special Purpose Processor Development Group where he performs signal and power integrity analysis.

**Richard B. Ericson** received a BSEE from Iowa State University in Ames, IA. He is currently a Project Engineer at the Mayo Clinic Special Purpose Processor Development Group where he focuses on signal and power integrity research and analysis.

**Clifton R. Haider** received a Bachelor of Science degree in Biomedical Engineering from the University of Iowa and a Ph.D. in Biomedical Engineering from the Mayo Clinic. He is currently Deputy Director of the Special Purpose Processor Development Group, directing research in high performance electronics and related areas.

**Barry K. Gilbert** received a BSEE from Purdue University (West Lafayette, IN) and a Ph.D. in physiology and biophysics from the University of Minnesota (Minneapolis, MN). He is currently Director of the Special Purpose Processor Development Group, directing research efforts in high performance electronics and related areas.

## Introduction

Integrated circuit (IC) components offering 56 Gb/s PAM-4 serializer/deserializer (SerDes) technology are beginning to become widely available. Design engineers are in the process of developing next generation SerDes technology – specifically for 112 Gb/s PAM-4 implementations.

An earlier paper [1] from 2016 demonstrated that printed circuit board (PCB) plated-through-hole (PTH) via technologies could support the 28 GHz Nyquist frequency for 56 Gb/s non-return to zero (NRZ) links. We demonstrated that 2-layer via stubs, roughly 10 mils long, were viable if padless vias are used – a technology that multiple board shops have no problem manufacturing. 112 Gb/s PAM-4 has the same baud rate as 56 Gb/s NRZ coding and thus has the same 28 GHz Nyquist frequency. Therefore, with restrictions such as lowering crosstalk to better resolve four voltage levels instead of two, 112 Gb/s PAM-4 should be viable using PTH PCB vias. Note that references to 28 GHz throughout this paper are to be interpreted as references to 56 Gbaud applications – specifically 56 Gb/s NRZ and 112 Gb/s PAM-4 signaling.

We selected a different vendor to manufacture a board that was supposedly identical to the PCB test board described in [1]. However, that second board, when tested, demonstrated much poorer via bandwidth. This result suggested that new design specifications needed to be developed to ensure high PTH via performance across vendors.

This paper begins by comparing electrical responses of the two vendors' PTH vias. Additionally, a new modeling methodology is proposed here to better assess the performance of PCB vias. The physical tolerances affecting via performance are also described. It is furthermore demonstrated herein how to apply physical tolerances in the PTH via modeling process and how to show the performance results using our new modeling methodology. This paper concludes with a proposed set of fabrication specifications that PCB vendors will have to meet such that measured PCB via bandwidth will meet expectations.

## Vendor Comparisons

The test board, described in considerably more detail by [1], has a cross-section shown in Figure 1. With modeling, it was determined that good PTH via bandwidth could be achieved with a via stub spanning up to two layers providing that the bottom pad is of "padless" design, that is, having a diameter approximately equal to the drill diameters. The board is comprised of a 12-layer sub-laminate above a 14-layer sub-laminate. PTH vias in the sub-laminates were then available as blind vias in the final board. In this approach, potential high-speed vias are available at routing layers 10, 12, 24, and 26. Layer-26 was a bottom layer and therefore not used due to component pad blockages. A final high-speed routing layer, not shown, was obtained using laser blind vias to Layer-3.

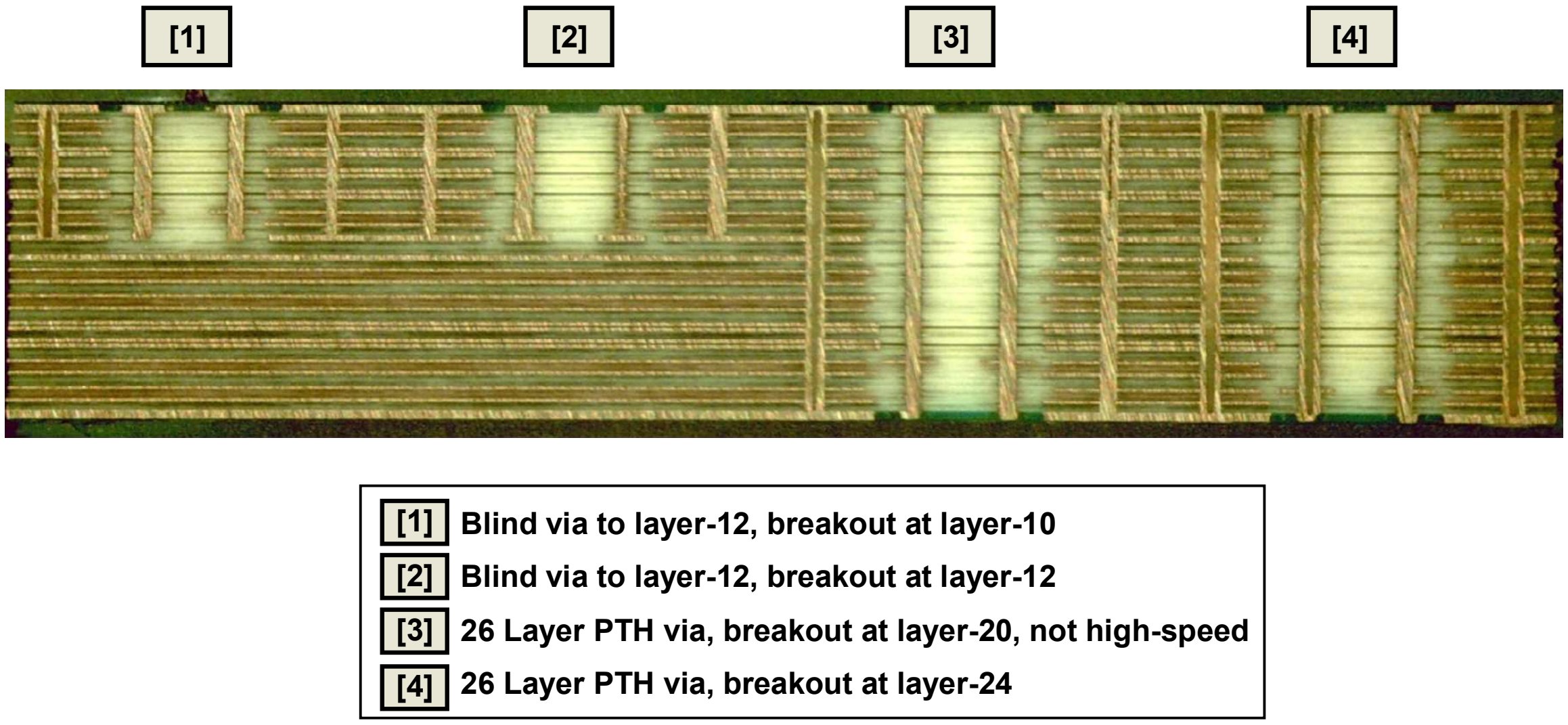

**Figure 1 – Test board cross-section showing high-speed PTH vias [46606]**

Excellent performance was measured with the first test board (built by “Vendor 1”). Next, the artwork was sent to “Vendor 2” who manufactured the second test board. The Vendor 2 board was tested and the measurements of 4.2 inch striplines (with accompanying via launches) were compared with those from Vendor 1, with the comparisons presented in Figure 2. The lower frequency SDD11 responses, all well below -20 dB, demonstrate that all stripline impedances are closely matched to 50 ohms. Therefore, the poor performance from the Vendor 2 board indicates some differences from Vendor 1 in the manufacturing of the via structures.

Figure 3 shows the SDD11 performance of each via type rather than the combined effects of via pairs connected by a long stripline. Measurements in this case are taken on minimal length striplines which are connected to top surface probe pads, using specific via technologies, to each high-speed stripline layer. Next, these measurements are ‘bisected’ [2] to show the electrical effects of a single isolated via/probe launch. Similarly, Figure 4 presents the time domain reflectometry (TDR) response for the four high-speed vias from both vendors’ boards. The TDRs demonstrate that the Vendor 2 boards have vias that, collectively, are more capacitive than those of Vendor 1.

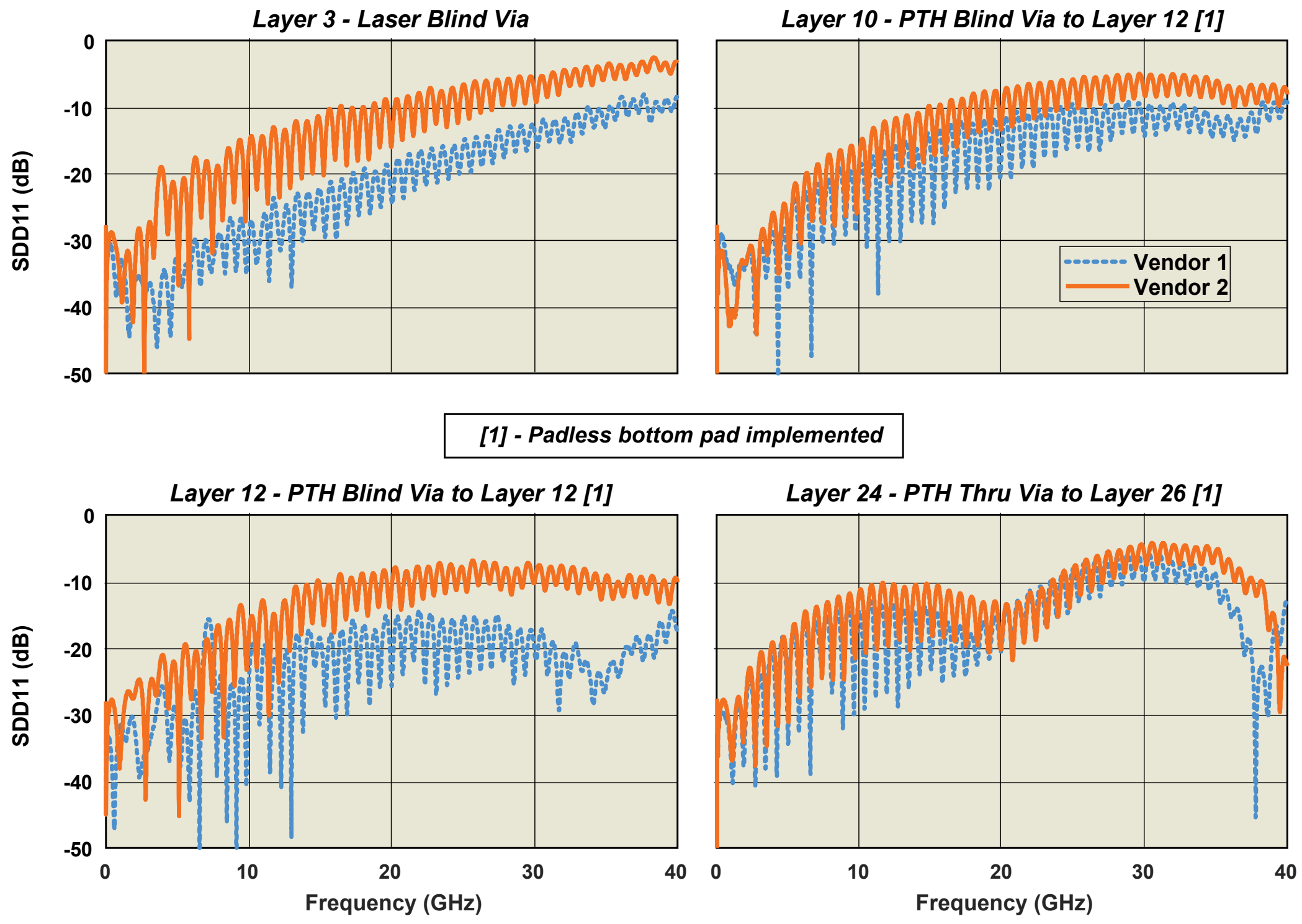

**Figure 2 – Vendor comparisons of SDD11 versus frequency for four 4.2 inch striplines each terminated with different via types [46617]**

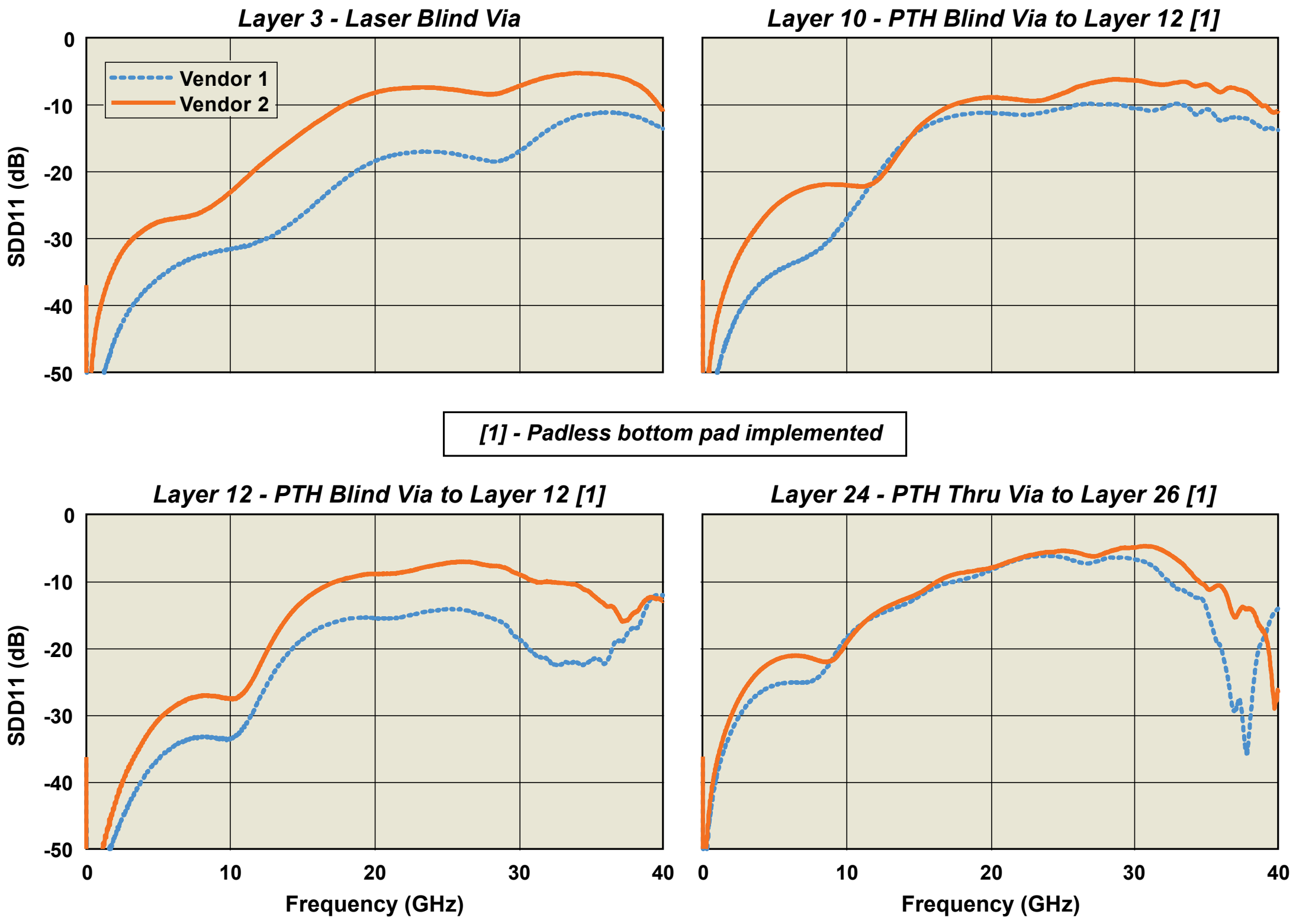

**Figure 3 - Vendor comparisons of SDD11 versus frequency of four via types [46608]**

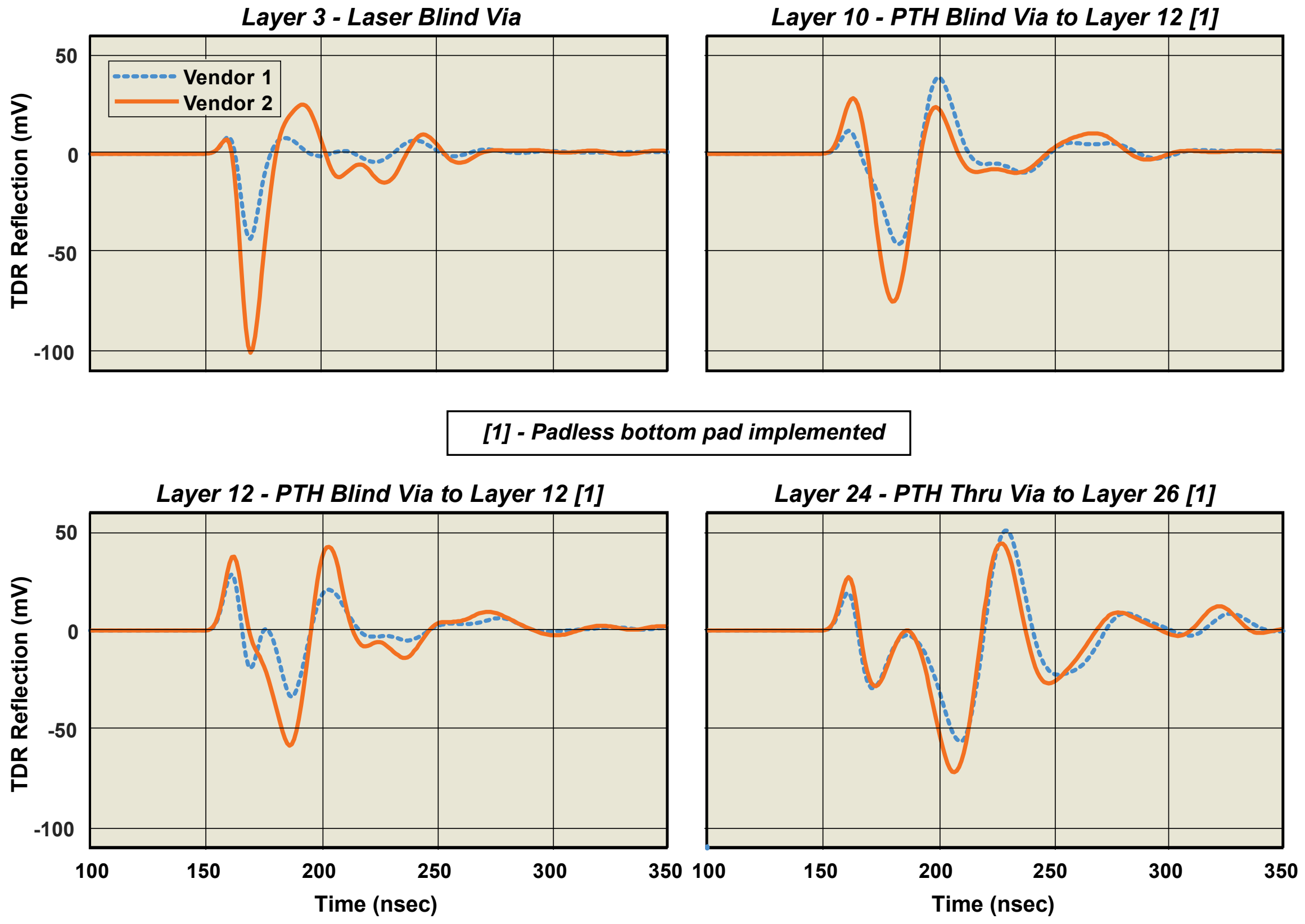

**Figure 4 - Vendor comparisons - TDR of four via types [46609]**

For an unknown reason, and not approved by us, Vendor 2 had enlarged the signal escape pads, as confirmed by vendor cross-section image comparisons on Figure 5 and Figure 6 for Layer-3 and Layer-10 break-out layers, respectively. These same oversized pads were noted for the Layer-12 and Layer-24 vias. The oversized pads are the primary reason for poorer measured electrical performance. Otherwise, anti-pad sizes, i.e., the copper plane voids around the via, appear to be similarly sized between the two vendors.

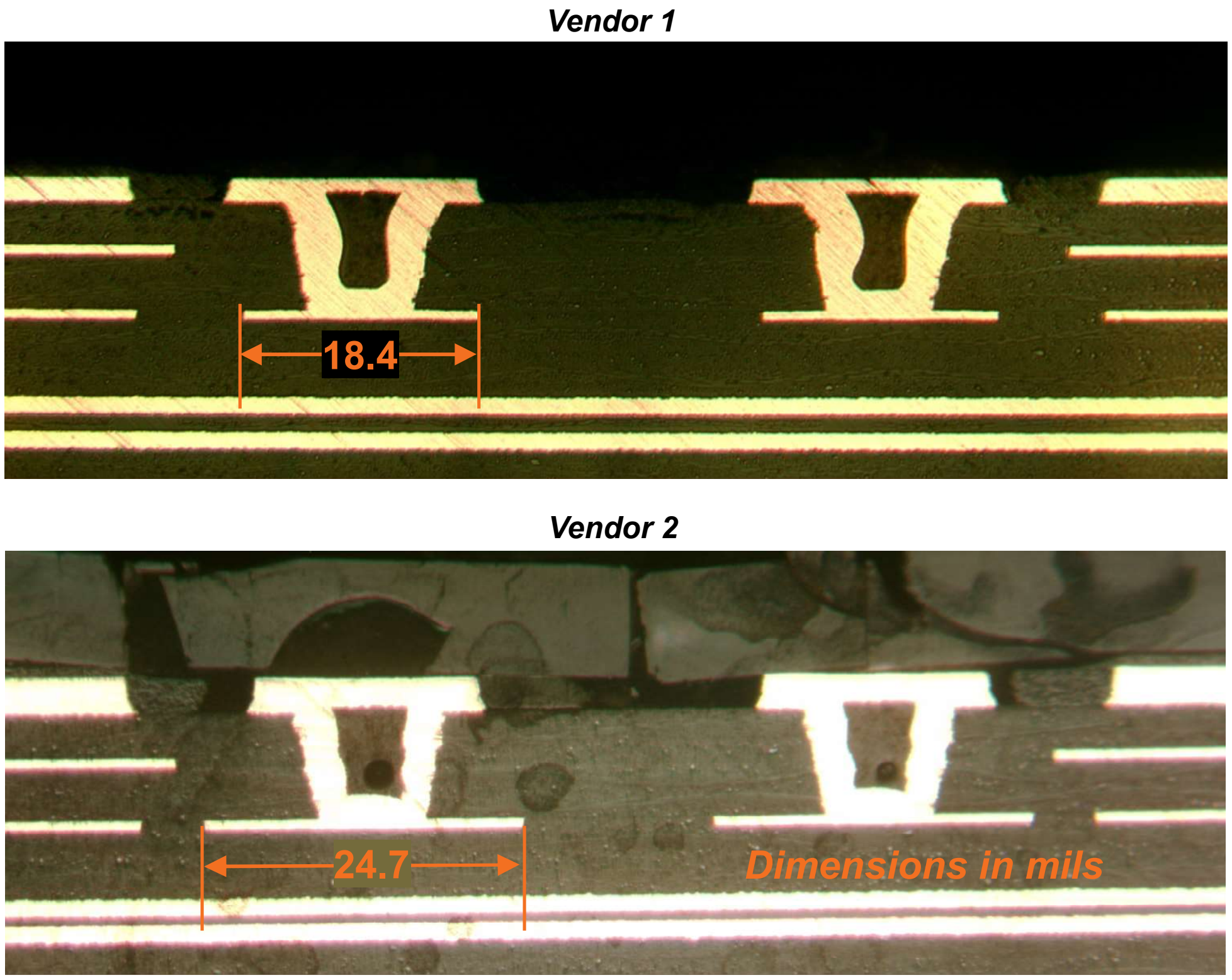

**Figure 5 – Vendor comparisons of Layer-3 via [46605]**

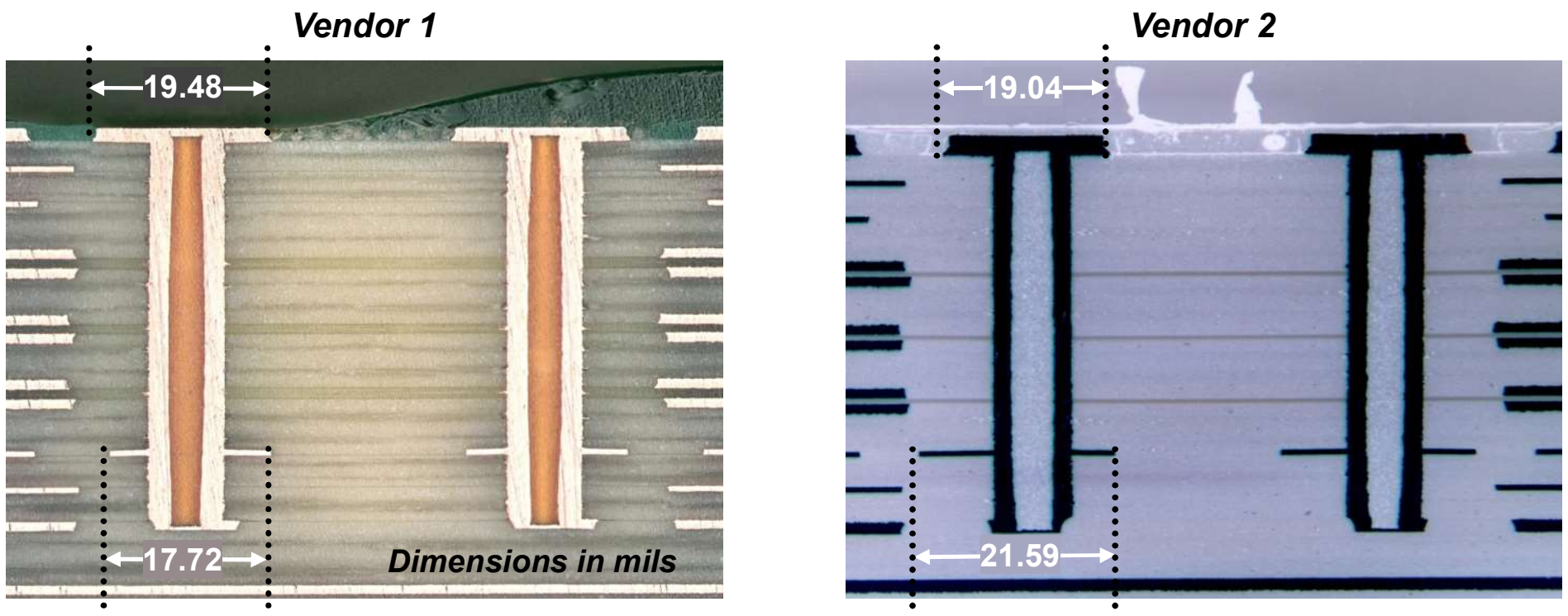

**Figure 6 - Vendor comparisons of Layer-10 via [466016]**

Clearly, PCB vendors need guidance beyond database design files for compliant via construction. Regardless of the design requirement communication method, a post-fabrication audit of the physical properties must also be considered. At 28 GHz, simple changes to via design or differing vendor manufacturing tolerances can cause via performance to not meet electrical requirements. But first, a robust and efficient approach is required to determine whether a via meets electrical requirements; this approach is the topic of the next section.

## Via Performance Characterization Methodology

Recently, we have adopted the practice of first establishing link performance curves (LPC) [3] before creating detailed 3-D via models. The LPC approach is revisited here because it employed a unique technique to determine the electrical bandwidth for a PCB via. Additionally, the LPC description below provides better context before our description of a proposed PTH via modeling methodology.

An LPC is presented in the lowermost portion of Figure 7 for the example link shown. The LPC independent X-axis is set to the "broadband reflected voltage", or BRV, as described in [3]. Stripline length, in terms of insertion loss, is added at each BRV point until the link begins to fail, thereby mapping out a curve where design points below the curve are predicted by simulation to meet passing criteria, be it eye opening or some other IC vendor-supplied criterion. The BRV attempts to assign a performance metric for a 3-D structure such as a PCB via which typically has a dynamic return loss over frequency. Briefly, a straight line is placed across the return loss "humps" and the BRV is taken in units of dB along this straight line at the Nyquist frequency. This method has proven to be robust but does require manual analysis. Alternatively, an automated process to access via performance is highly desired. Specifically, at 28 GHz, it is expected to be difficult to design a PTH via using worst case design parameters (corner cases) and still meet electrical performance goals. In practice, it is extremely unlikely that all worst-case design parameters can be achieved. With a Monte Carlo analysis, one simply creates N models, randomly picking from the design parameter statistical distributions. Then the results of the N models can be statistically analyzed to determine the variation of the electrical performance. Therefore, with large N, a uniform and automated approach is needed to determine via performance.

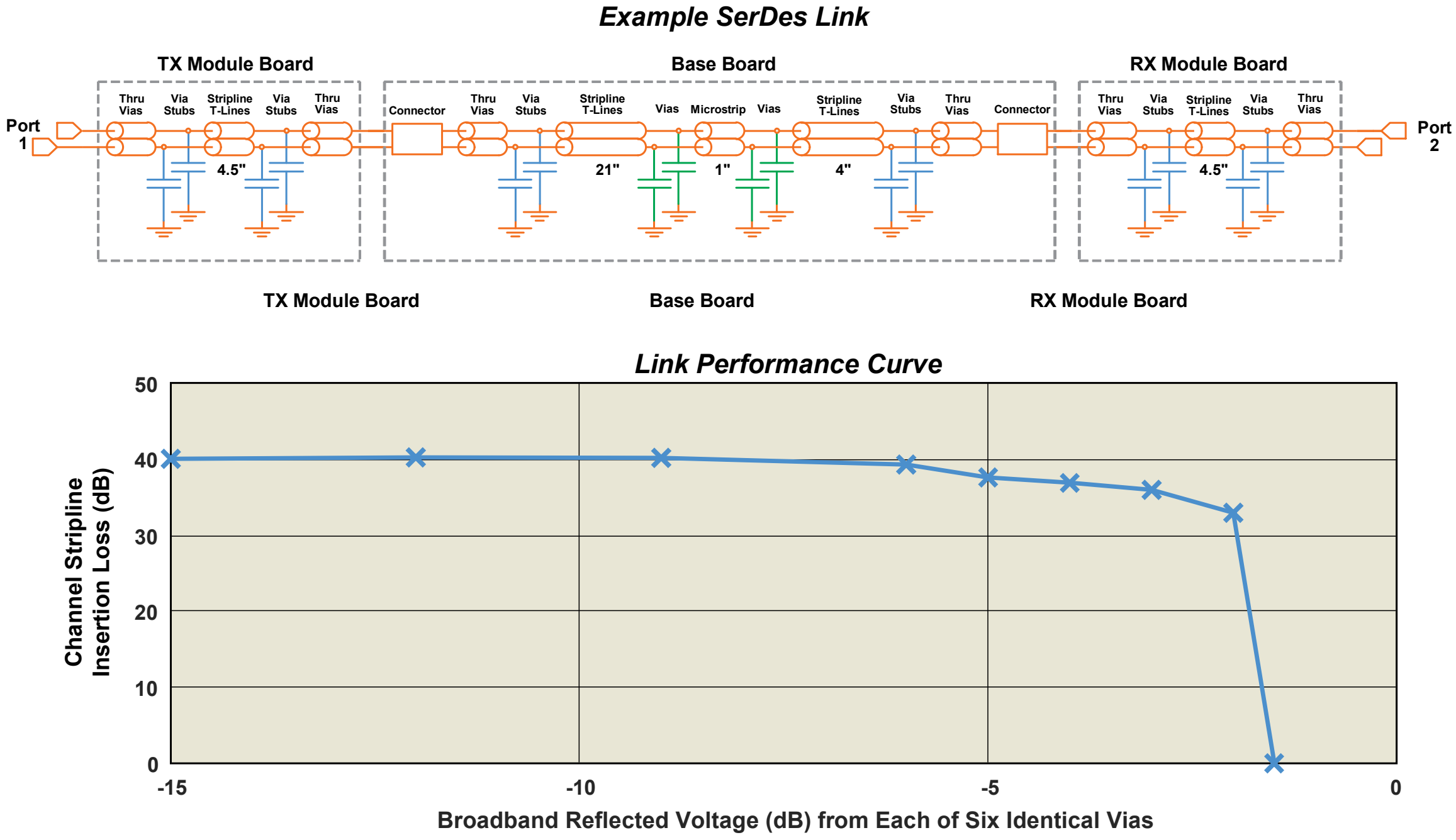

**Figure 7 – Example SerDes link and associated link performance curve [46604]**

A first attempt to automate the via performance was to calculate a pulse response from the modeled S-parameters - in this case, an 18 picosecond-wide pulse corresponding to a 56 Gb/s symbol. Rising and falling pulse edges were chosen to be sinusoidal to limit high frequency harmonic content. The same pulse/edge shape will be used in further analyses discussed through the remainder of this paper. To serve as a test case, the Layer-10 break-out via was modeled with six different bottom pad diameters ranging from 0.2510 mm (padless) to 0.4550 mm (full-pad) diameters. The results of this analysis appear in Figure 8. Unfortunately, the pulse response did not differ much even though a significant degradation in electrical performance was expected with the increasing bottom pad size. However, a significant difference can be observed, as plotted in Figure 9 in the voltage reflected from a pulse response. Reflected behavior is not always a good metric for determining performance as it pertains to a SerDes channel. For example, short lengths of higher or lower impedance striplines are typically well tolerated in a SerDes link yet can reflect significant voltage in response to a voltage pulse. It should be noted that these results are achieved using internally generated MATLAB® scripts. Keysight's Advanced Design System (ADS) was alternatively used to create pulse and reflected pulse responses. ADS did predict greater pulse response differences, but the reflected pulse showed significant ringing. Additionally, ADS offered several different interpolation settings comprising 48 unique combinations, some of which gave significantly differing results. Thus, our MATLAB scripts were used for our analysis since consistent results could be achieved.

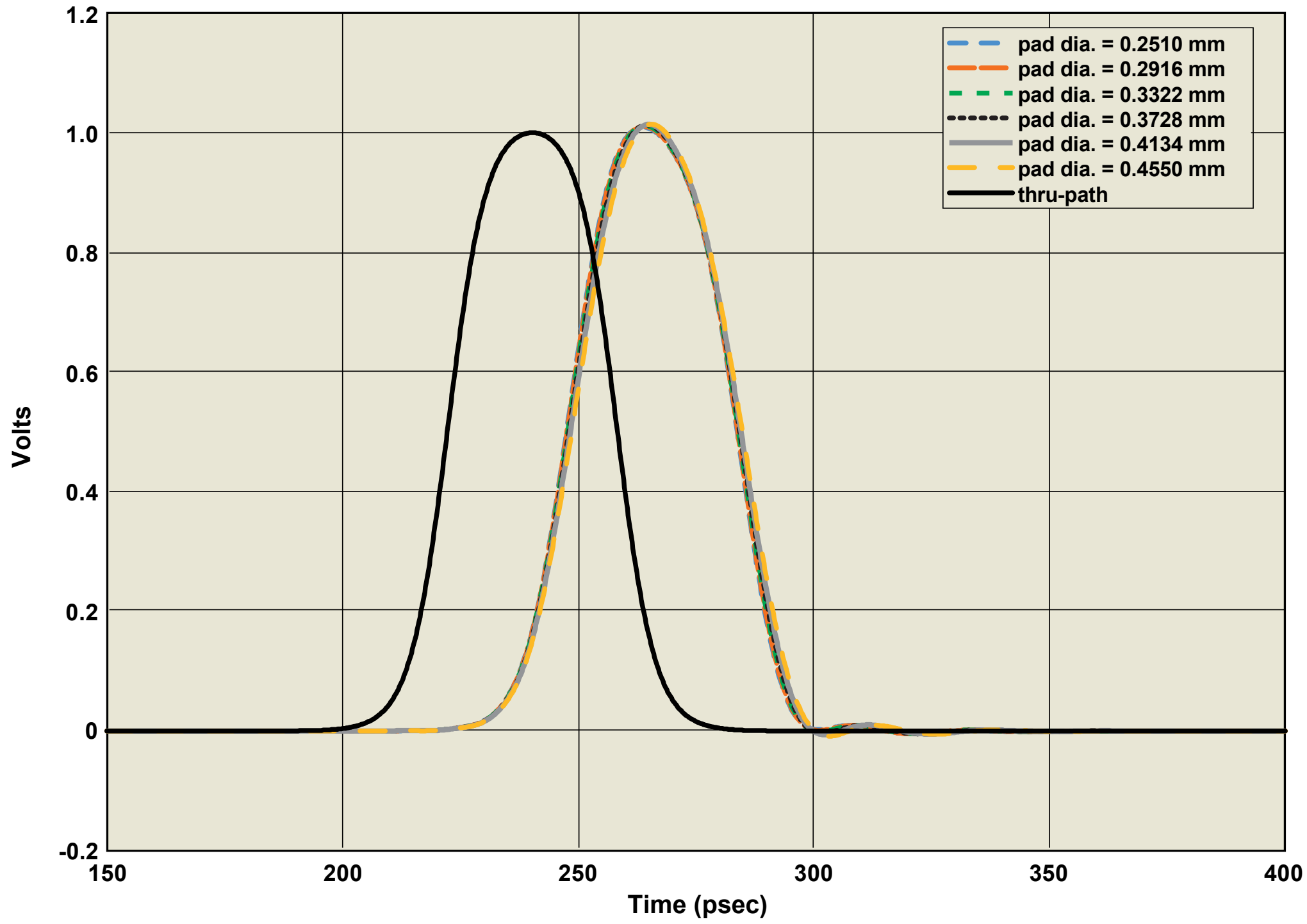

**Figure 8 – Pulse response for 12-layer via with Layer-10 break-out – differing bottom pad sizes [44610]**

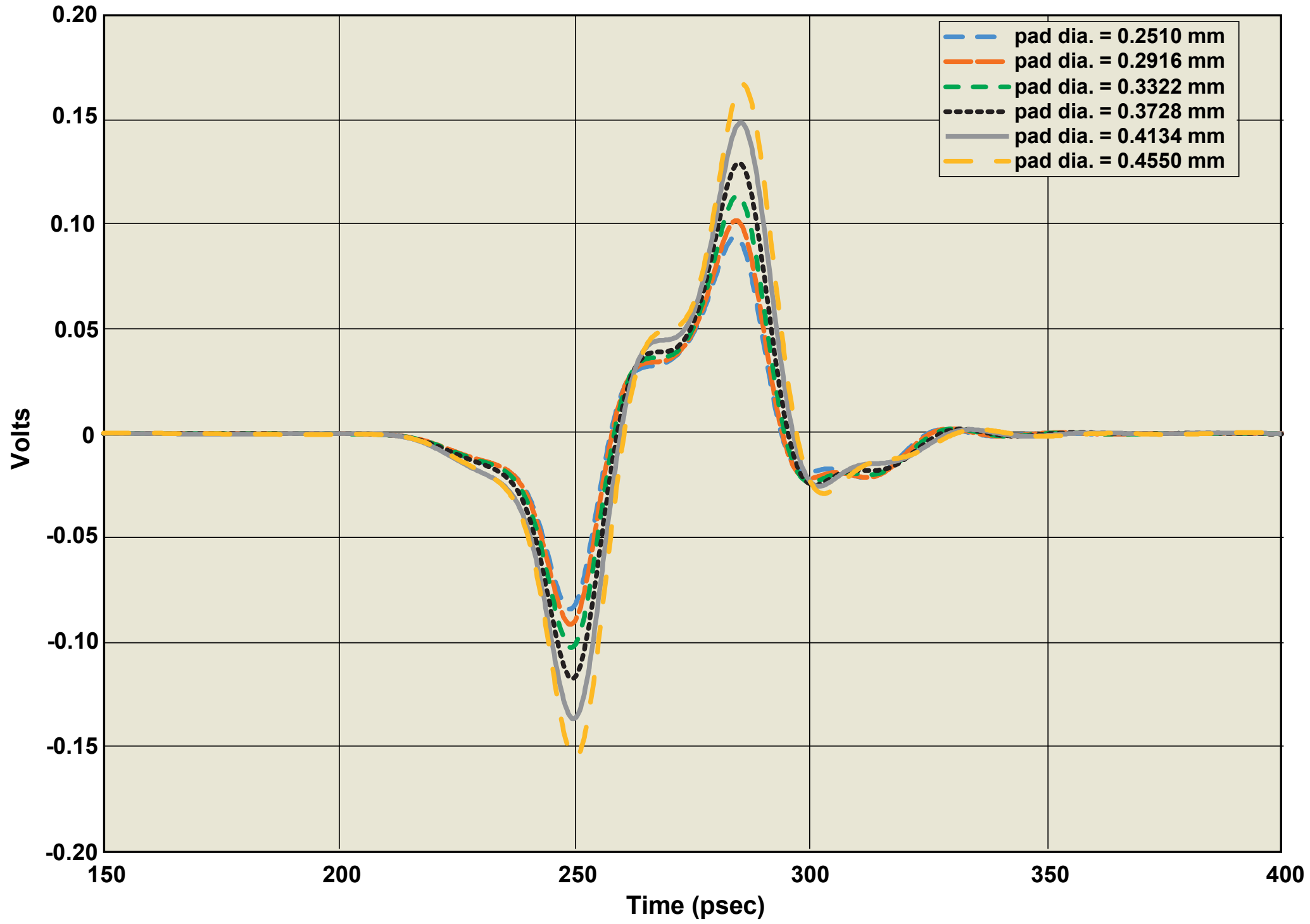

**Figure 9– Reflections from pulse response for a 12 layer via with Layer-10 break-out – differing bottom pad sizes [44611]**

Ultimately, eye closure was selected as the new automated metric for via performance. In this method, eye closure due to via model distortions is initially compared to an undisturbed eye diagram. A 2^7-1 PRBS pattern was chosen for the stimulus for expediency. Longer patterns were not thought to be needed because the transient behavior through the short vias should dissipate quickly. It should be noted that lossless dielectric and metal materials were used to minimize via loss. The initial thinking was that material loss may be corrected by link equalization. The use of the lossless materials does not prevent the via model from exhibiting plane-mode insertion losses discussed in [4] as we did not want to neglect these losses. Using a repeating bit pattern as a source has a finite spectrum that is discrete. Thus, the mathematical approach to create eye diagrams can be direct without the need extending the bandwidth to accommodate higher harmonics. The single-pulse response has a much wider spectrum which must be convolved with the finite bandwidth channel model – or partial channel model in this case. In doing so, assumptions have to be made and heuristically, the PRBS/eye-diagrams appear to give better results when compared to the single pulse approach.

Frequency responses and inner eye outlines for the via model described above are plotted in Figure 10. Also plotted are results for lumped capacitors having values computed using the BRV approach (i.e., where the capacitor value is set to match the BRV value at 28 GHz). The eye closures using the 3D models are clearly much worse when compared to those using the lumped capacitor. The 3D model SDD11 responses above 28 GHz are generally many decibels higher than that of the corresponding lumped capacitors, which offer a single pole increase in SDD11.

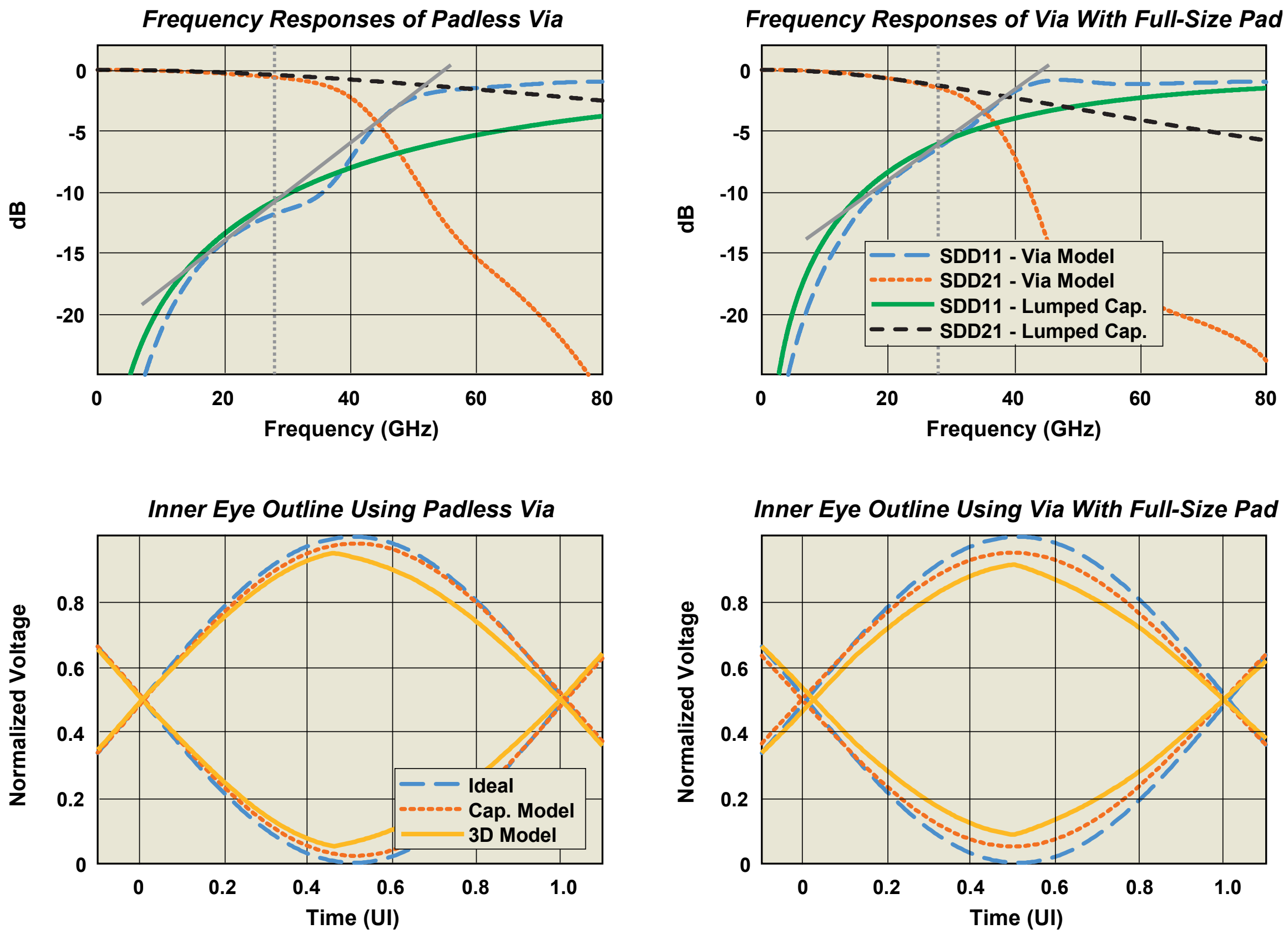

**Figure 10 – Frequency response and eye closures of via models versus lumped capacitors [46613]**

The PRBS pattern exhibits high frequency spectra that are substantially attenuated by a lossy channel. Therefore, eye closure attributed to the PTH vias may be quite different when considering single discrete structures versus structures used as part of a lossy channel. To study this further, PTH via models were compared in a lossy channel that were intended to replicate the electrical behavior from channels, whose measurements are displayed in Figure 2. Specifically, three channels were examined using via technologies of 1) Vendor 1 laser blind vias, 2) Vendor 2 laser blind vias, and 3) Vendor 1 Layer-10 break-out vias. Next these three models were compared for eye closure as isolated models and then as part of a lossy channel. Simplistically, the lossy channel was formed by chaining via models to the end of 0.1 meter stripline models having 15 dB of insertion loss at 28 GHz. By the BRV approach (and other conceivable approaches), it was expected that the Layer-10 break-out via should have eye closure somewhere between that of the two Layer-3 blind vias. However, with the earlier observations, the Layer-10 break-out via exhibited worse eye closure than the Vendor 2 blind via model when studied as an isolated structure, as illustrated in the upper right sub-panel of Figure 11. Using the via models in the lossy channel indicates that eye closure reported in the lower right sub-plot appears to track with the BRV at 28 GHz. This result is fortuitous in that PTH technology may produce a very high return loss above 28 GHz, much more so than the single-pole return loss from a lumped capacitor; yet PTH vias are still viable for 28 GHz applications. Hence, eye closure due to a PTH via in a lossy channel will be used, as opposed to the BRV, for the measure of the via bandwidth.

For completeness, it should be noted that the lossy channels completely closed the eye diagrams without any equalization applied.   Therefore, these eye diagrams were equalized with a 2-tap FIR filter which opened the eye diagram to about one-third that of the 1-volt input voltage.  Since the attenuation of the channel reduces the eye closure from PTH vias for 56 Gbaud applications, then ideally the eye closure should be benchmarked in full link simulations using IC vendor transceiver models and channel models that match expected hardware implementations.

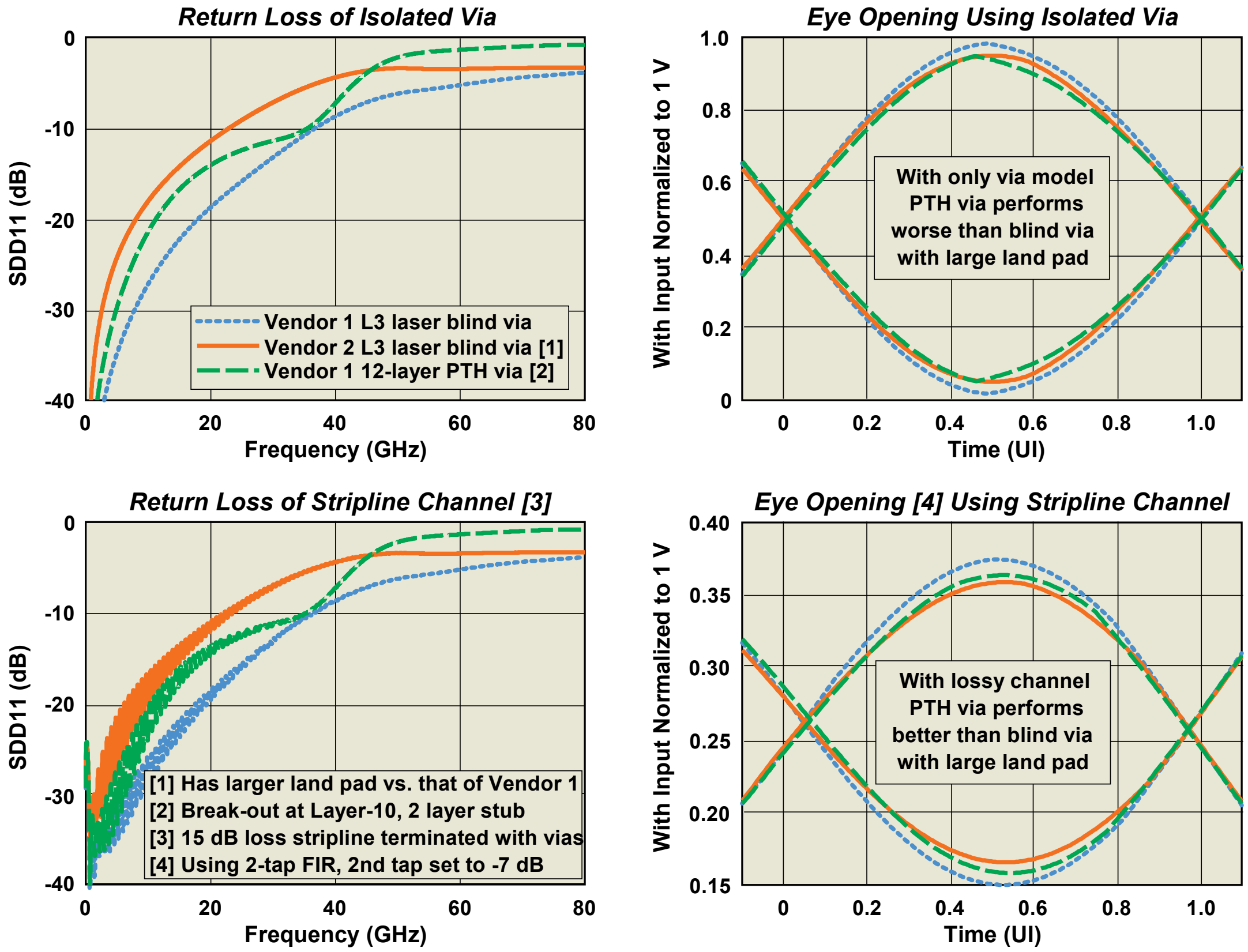

**Figure 11 – Comparing eye closures of 3D models – isolated model versus lossy channel [46623]**

With eye closure in equalized lossy channels selected as the performance metric for PTH vias, the next step is to identify physical and electrical variability for the purposes of accurate PTH via modeling.

## Effective Via Barrel Diameter

PTH vias are first drilled with mechanical bits.  Therefore, it is not expected that the drilled hole diameter has much variability.  First described in [4] and presented in Figure 12, during plating of the via barrel, some copper actually becomes plated, or “wicked”, along the glass reinforcement fibers. The wicking among three via barrels is shown along with a close-up of the effect on the middle barrel.  Little is known about how commonplace the wicking occurs, but it has been present in several boards that we have cross-sectioned (where we have observed wicking reaching into the sidewalls of the via barrel approximately 2 mils). We note that the images in Figure 12 are from a different PCB with worst case wicking to provide a clear visualization of a potential fabrication issue.

Much of the electrical field fringing from the barrel structure should originate on the wicked fibers. Regarding the vendor boards studied in this paper, based on Figure 6, the via wicking effectively grows the via barrel outward by approximately 1 mil, thereby increasing the via barrel outer diameter by 2 mils. This effect is another variable to account for when considering via electrical performance.

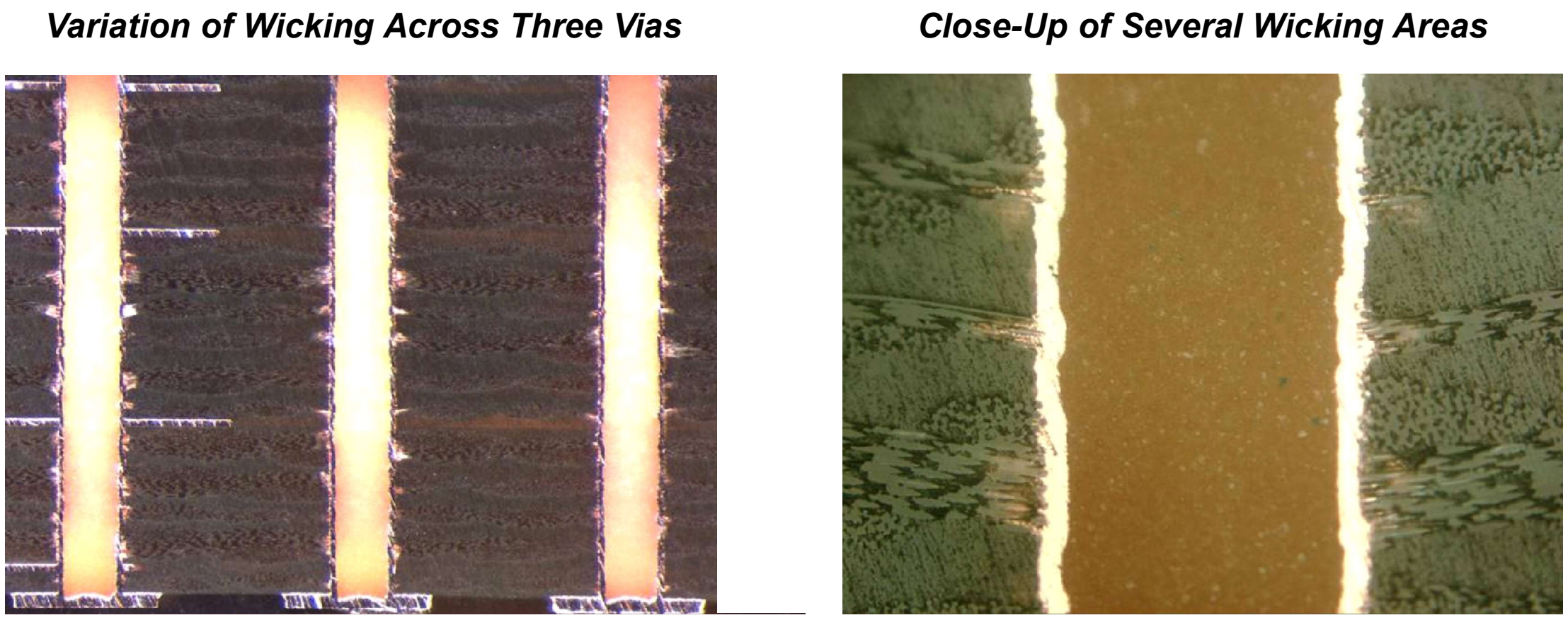

**Figure 12 – Examples of copper wicking along fiber bundles at PTH barrel boundary [46614]**

## PCB Layer Shifting

At very high frequencies, one should not assume that all PCB layers are aligned when modeling the electrical performance of PTH vias. In [1] a method was described using X-ray imaging to measure layer alignment within a PCB. A simpler method employed by PCB fabricators is to incorporate a "plus" pattern of metal replicated and aligned on all metal layers. After fabrication, the center of the pattern is drilled with a tapered bit and measurements of the copper strip locations can be used to calculate layer registration. An example pattern, where the tapered drilling has been implemented, and a plot of measured layer shifting appear in Figure 13. PCB vendors can often adhere to 2 mil layer shifting (from true position) even for complex PCB boards with more than 20 layers.

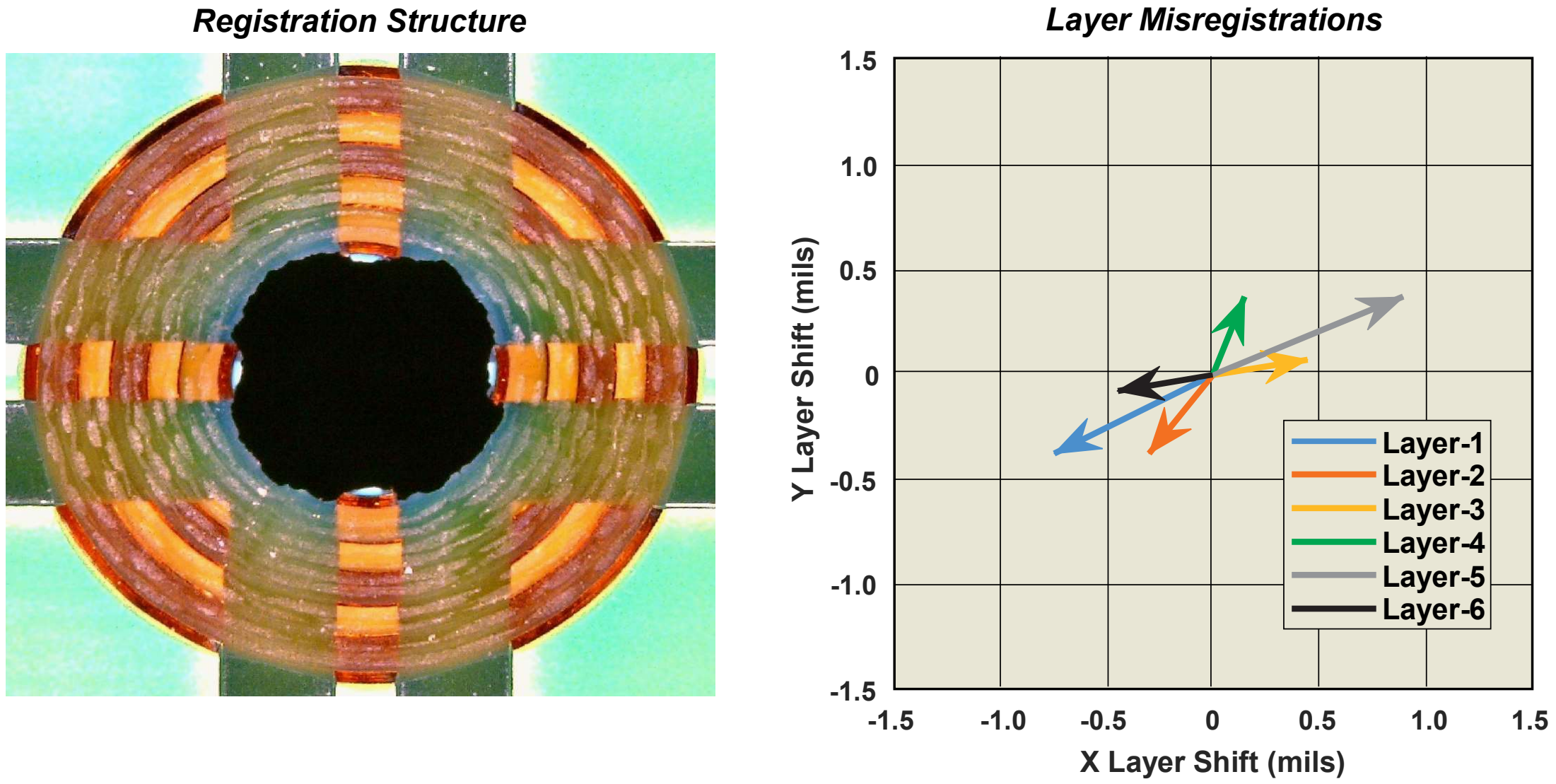

**Figure 13 – Structure to measure PCB layer to layer alignment [46615]**

## Comprehensive Modeling

As discussed above, PTH via models can have dimensional variability due to layer shifting and effective barrel outer diameter variability due to wicking. Additionally, copper etch tolerance is well known in regard to impedance control, but it also will affect finished antipad sizes. These three factors were used to build sets of models with dimensional variability. Several other factors were not considered, such as via drill tolerance and dielectric variability. For better modeling accuracy, it is important to include the dielectric anisotropic behavior as discussed in [4,5].

The model for the Layer-12 blind via with Layer-10 breakout was analyzed using 99 Monte Carlo cases with the following physical property variations:

- 0-2 mil layer shifting
- 8-10 mil outer via diameter
- Line width variation allowing 100 +/- 7 Ohm differential impedance

Uniform distributions were selected for all three varying parameters. Via anti-pad sizes were set to track with line width tolerance, since all structures are created during the same copper etch process. However, the percentage change in the anti-pad sizes due to etch variation is much smaller than with stripline width variation. Therefore, anti-pad size variation should not have much effect on the distributed (characteristic) impedance of the via barrel. Other distribution types, such as Gaussian, would result in less overall parameter variation but a more conservative analysis (using uniform distributions) was chosen. It is generally assumed that PCB vendors screen their boards to verify characteristics such as line width and layer registration specifications, therefore scrapping cores and entire boards is not that uncommon. Therefore, assuming Gaussian distributions could be overly optimistic.

Figure 14 illustrates angled and top views of one of the models used for Monte Carlo simulations. The layer shifting is more apparent in the top view.

***Angled View of Model***

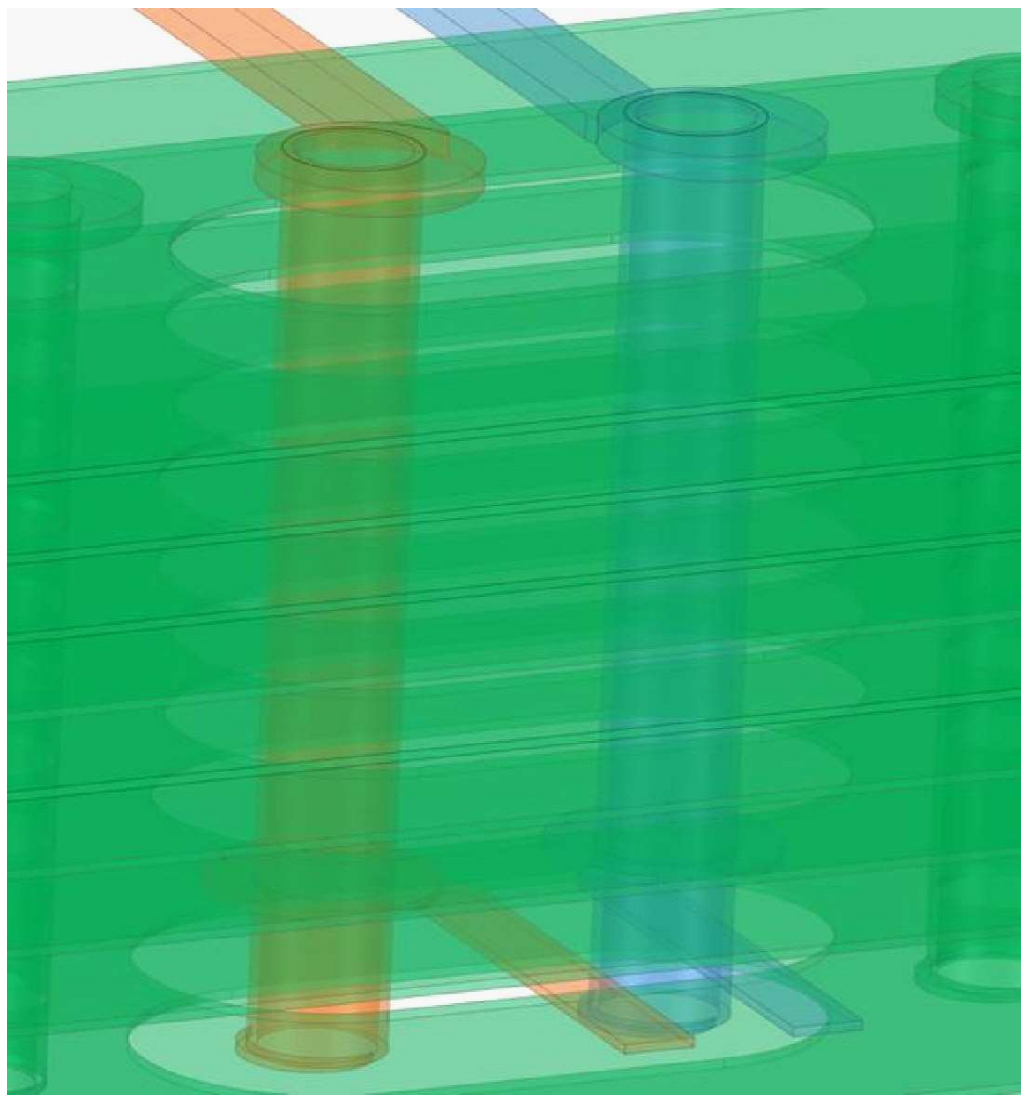

***Top View of Model Showing Layer Shifting***

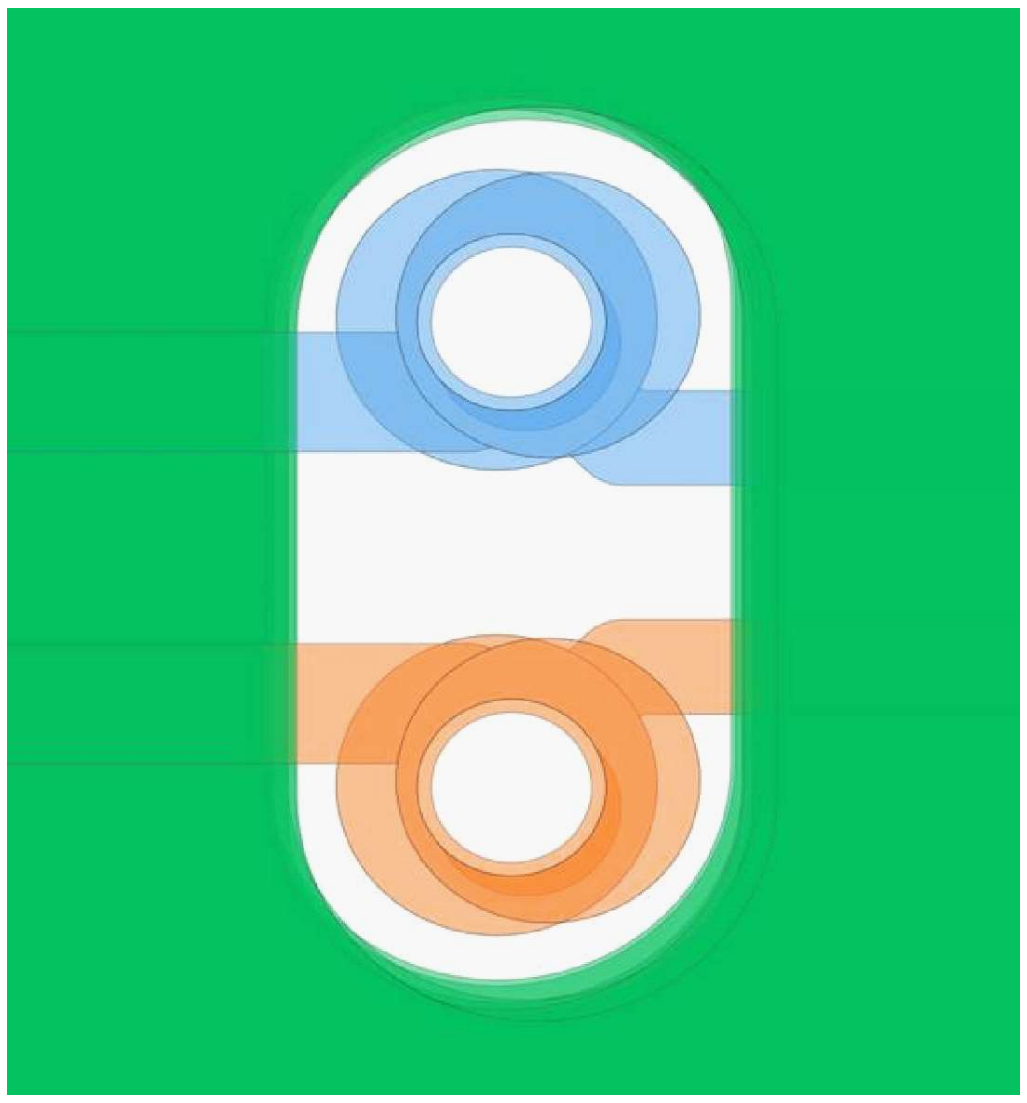

**Figure 14 – One of 99 HFSS models illustrating manufacturing variability [46622]**

A histogram binning of the extracted eye closures appears in Figure 15. The standard deviation and mean were calculated from the data set and projected that the 3-sigma eye closure limit would be just below our 0.05 mV*UI acceptance criterion. The eye closure limit was arbitrarily set to illustrate the design flow, and in practice acceptable eye closure would require a link budgeting study. Furthermore, it is not known whether 3-sigma compliance is required. There are many other parts of the link, such as TX/RX circuits and stripline loss variability, that if included in the link analysis, along with their statistical distributions (rather than using worst corner cases), would decrease the overall link performance variability. Note that the eye closure is plotted at 'nominal case', which is defined to be the eye closure with no layer shifting, 100-ohm trace impedance, and 8 mil via diameter. It is of interest that the histogram is well centered about this nominal case. We expected that the nominal case would provide the lower limit to the eye closure because the layer shifting and via barrel diameter above 8 mils should have resulted in an overly low via impedance. These results suggest that this PTH via can be redesigned for better electrical performance.

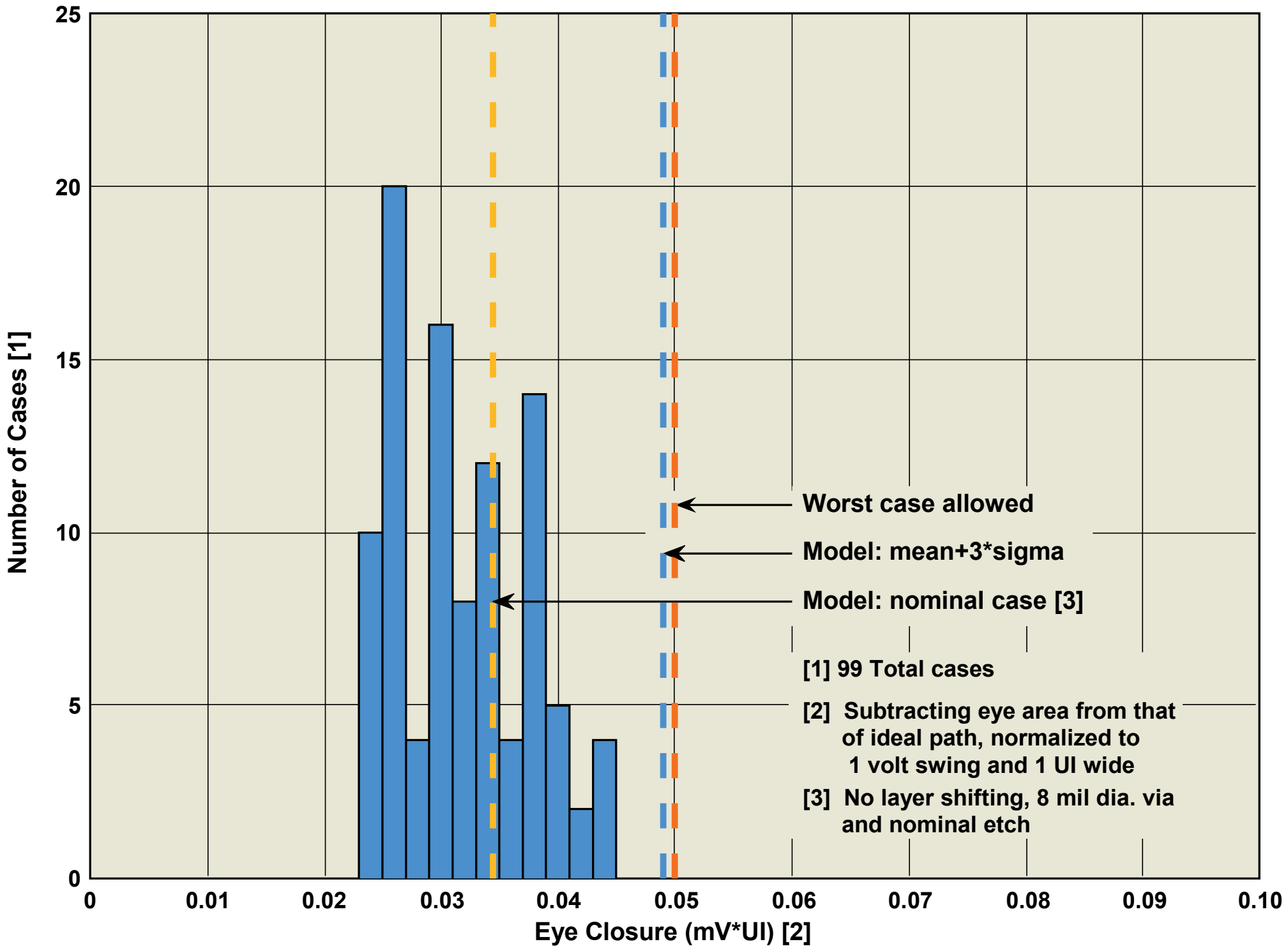

**Figure 15 – Monte Carlo simulations of PTH via allowing 2 mil layer shifting [46624]**

As commented previously, it would be better to determine eye closures with a link simulator using vendor-supplied TX/RX models and channel models projected to match hardware implementations. However, it might be very difficult to simulate the 99 cases that were used to generate each histogram. It may be worthwhile to determine whether using simpler channels and minimal equalizing approaches would yield comparable eye closure behavior obtained with more comprehensive link modeling. If so, then simulations could be conducted using MATLAB, which was used to determine eye closure. The MATLAB script employed to read in Touchstone® files and calculate the eye closure required approximately 30 seconds, with much of that time used to read the Touchstone files from disk. It should be noted that to create the Touchstone files, approximately 24 hours of run time were required to simulate the 99 via models with Ansys HFSS.

The Monte Carlo simulations were repeated with the layer shifting increased from 2 to 3 mils. This study was originally intended to demonstrate that the PTH via performance was very sensitive to layer shifting. However, the results demonstrated that the statistical worst case, as illustrated in Figure 16, only increased the eye closure slightly, just beyond the 0.05 mv*UI worst-case target. These results are intriguing as they seem to support that PTH via performance is fairly robust to layer shifting. Perhaps a similar study with a longer PTH via may impose increased performance degradation in response to 3 mil layer shifting.

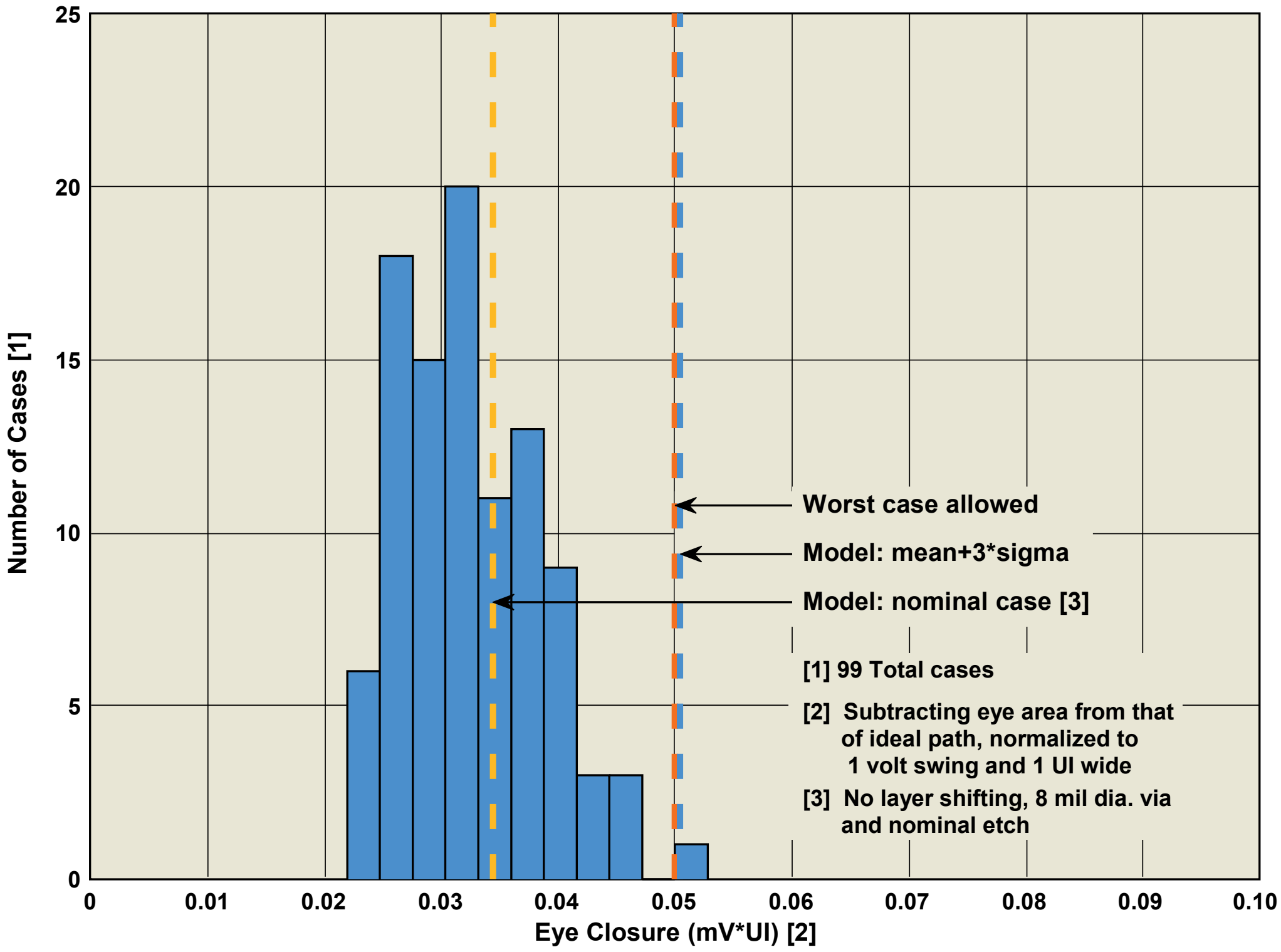

**Figure 16 - Monte Carlo simulations of PTH via allowing 3 mil layer shifting [46625]**

## Rule Development

PCB designers typically include a set of rules on a fabrication drawing (frequently referred to as "fab notes") that they send to the board fabricator, along with the PCB design database. A subset of these rules is specifically intended to ensure high performance electrical conformance. Two common high-speed rules are: 1) setting the characteristic impedance nominal and tolerance values, and 2) mandating artwork rotation to reduce fiber-weave-skew. To this end, we suggest the following example additional rules to enforce electrical performance specifically for laser and PTH vias:

1) Via pad diameters shall not be changed from those as designed.
2) Layer-to-layer shifting shall not exceed 2 mils from center.
3) Copper wicking along glass fibers during via PTH process shall not exceed 1 mil in depth.

In some cases, more stringent rules may be necessary, depending on differing board thicknesses and myriad other build variations. For best results, we recommend following an analysis approach similar to that previously described in this paper, to develop a set of rules tailored to specific requirements.

One may wish to discuss these requirements with the PCB vendor to ensure that requirements are adequately conveyed by the new rules. To add another level of assurance, it may be advantageous to consider having the vendor measure features in early builds ("scout lots") to verify compliance.

For the PCB described in this report, we noted two other PTH via types that demonstrated 28 GHz capability. Those vias would also need the same modeling process described above to determine if the suggested rules are adequate. In the future, the hope is also to study the performance impacts of drill tolerance positioning, which in turn may result in a fourth rule specifying drill accuracies for the vias used for high-speed signaling. The present understanding is that layer-to-layer shifting control limits may already be commonly adopted by PCB vendors. Note that PCB vendors are usually limited to testing etched line impedances, and few have the capability to measure the electrical performance of vias for 28 GHz applications. Therefore, the customer (the signal integrity engineer) will be responsible for identifying mechanical rules for the PCB vendor that result in electrical performance compliance.

## Summary

We have developed an approach to guarantee electrical performance levels for PCB via technologies in response to finding a refabricated known-good board underperforming when compared to the original board. In this case, the performance degradation was determined to be attributed to the second vendor increasing pad sizes from the design database. However, this experience prompted the development of an approach to help ensure that PCB vias meet their electrical performance goals.

As a result, three fabrication specifications are proposed in this paper to be added to the fabrication drawing for use by the PCB vendor to meet customer requirements. One specification restricts the vendor from making changes to the pad sizes. A second states that the allowable layer-to-layer registration be within 2 mils. The third restricts the copper wicking depth to be less than 1 mil away from the PTH via barrel. Additionally, a fourth specification is proposed to stipulate the PTH via drill placement tolerance; however, via drill accuracy impacts on electrical performance have not yet been investigated.

A process was developed to determine if dimensional tolerances for the design of the PTH via were acceptable. First, HFSS was used to model 99 Monte Carlo cases with dimensional distributions set to the limits established in the fabrication note specifications. Then the Touchstone files from the HFSS simulations were used as input to MATLAB scripts used to predict eye closure. Further study was recommended in determining the detail required in the channel modeling in order to best predict eye closure. Also, an analysis will be required to determine where to set the eye closure limit, since the value was arbitrarily chosen to illustrate the methodologies described in this paper.

This study concentrated on just one PTH via definition. Simulation results for this particular via suggested that better choices for nominal values may result in increased electrical performance. Additionally, all proposed high-speed via definitions would need to be reviewed to verify the proposed specifications.

This paper excluded an analysis of crosstalk behavior for 56 Gbaud links. Assuming that crosstalk levels are acceptable, then the PTH via does seem to be viable for 56 Gbaud applications including 112 Gb/s PAM-4 links.